\documentclass[aps, prb, twocolumn, noshowpacs, noshowkeys, notitlepage, twoside, a4paper]{revtex4-1}
\usepackage[usenames,dvipsnames]{color}
\usepackage{graphicx}
\usepackage{amsmath}

\begin{document}
\title{SQUID ratchet: Statistics of transitions in dynamical localization}
\author{Jakub Spiechowicz}
\affiliation{Institute of Physics, University of Silesia,  41-500
Chorz{\'o}w, Poland}
\affiliation{Silesian Center for Education and Interdisciplinary Research, University of Silesia, 41-500 Chorz{\'o}w, Poland}
\author{Jerzy {\L}uczka}
\affiliation{Institute of Physics, University of Silesia,  41-500
Chorz{\'o}w,  Poland}
\affiliation{Silesian Center for Education and Interdisciplinary Research, University of Silesia, 41-500 Chorz{\'o}w, Poland}
\email[]{jerzy.luczka@us.edu.pl}
\begin{abstract}
We study occupation of certain regions of  phase space of an asymmetric superconducting quantum interference device (SQUID) driven by thermal noise, subjected to an external ac current and threaded by a constant magnetic flux.
Thermally activated transitions between the states which reflect three deterministic attractors are analyzed in the regime of   the \emph{noise induced dynamical localization} of the Josephson phase velocity,  i.e. there is a temperature interval in  which the conditional probability of the voltage to remain in one of the states is very close to one. Implications of this phenomenon on the dc voltage drop across the SQUID are discussed. We detect the emergence of the power law tails in a residence time probability distribution of the Josephson phase velocity and discuss the role of symmetry breaking in dynamical localization induced by thermal noise. This phenomenon illustrates how deterministic-like behaviour may be extracted from randomness by stochasticity itself. 
\end{abstract}
\maketitle
\section{Introduction}

Determinism is often related to \emph{predictability}, i.e. the ability to predict the future state of a system given the present one. Perfect predictability obviously implies strict determinism but lack of predictability does not necessarily means lack of determinism. The latter is often identified with \emph{randomness}. 
Nowadays we know that randomness understood as lack of predictability may emerge in deterministic systems which are chaotic \cite{ott2002}. In this work we would like to discuss a question concerning the reverse action, i.e. whether deterministic-like behaviour can be extracted out of randomness or at least randomness can be significantly suppressed by stochasticity itself. 
\begin{figure}[t]
	\centering
	\includegraphics[width=0.85\linewidth]{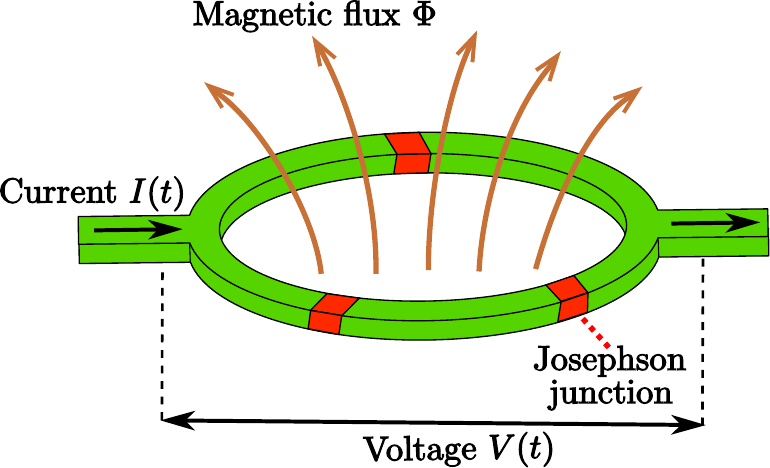}
	\caption{The asymmetric SQUID device built of the three identical Josephson junctions. It is driven by the external current $I(t)$ and pierced by the magnetic flux $\Phi$. The voltage drop across the SQUID is $V(t)$.}
	\label{fig1}
\end{figure}

To attack this problem we consider a SQUID \cite{zapata1996, sterck2005, sterck2009, spiechowicz2014prb, spiechowicz2015njp, spiechowicz2015chaos}. 
It offers some important advantages over other potentially useful setups: (i) precise experimental control of applied driving forces  in the form of external currents, (ii)  detection of directed motion manifested in a non-zero long-time dc voltage, (iii) access to studies over a wide frequency range of adiabatic and non-adiabatic external perturbations and finally (iv) both underdamped and overdamped dynamics can be investigated by proper junction fabrication and variation of system parameters. We want to emphasize that our findings apply to a broad selection of physical systems and could be experimentally observed in variety of setups including in particular type II superconducting devices based on motion of Abrikosov vortices \cite{lee1999,villegas2003}, Josephson vortices \cite{ustinov2004, beck2005, knufinke2012}, cold atoms in optical lattices \cite{renzoni2003, renzoni2005, lutz2013, arzola2011, arzola2013} and many others \cite{hernandez2004, costache2010, drexler2013, serreli2007, roche2015, grossert2016, kedem2017pnas, kedem2017nano, kedem2017, hongru2017}.

The layout of the present paper is organized as follows. In section II we briefly describe a model of the asymmetric SQUID driven by the external current and pierced by the constant magnetic flux. We also discuss deterministic dynamics of the system in the parameter regime considered in the remaining part of the paper. In section III we present main results. In particular, we first turn to its noisy counterpart to demonstrate the emergence of  dynamical localization of the Josephson phase velocity induced by thermal noise. Next, implications of this phenomenon on the dc voltage drop across the SQUID are discussed. We study there also probability distributions for the residence time of certain regions of the phase space of the model. Finally, this section includes a discussion on the role of symmetry breaking in the observed noise induced dynamical localization. Last but not least, section IV provides summary and conclusions.
\section{Model}
For the paper to be self-contained, we present a brief description of the SQUID ratchet, see Fig. \ref{fig1}. We refer the interested reader to Ref. [\onlinecite{spiechowicz2014prb}] for a more detailed description. This asymmetric device is composed with the three identical resistively and capacitively shunted Josephson junctions: two are placed in one arm whereas the third is located in the other arm. The system is pierced by an external constant magnetic flux $\Phi$ and driven by a current $I(t)$. The Josephson phases $\varphi_u$ and $\varphi_d$ are across junctions which are placed in the same arm. The dimensionless  Langevin equation governing the dynamics of the Josephson phase has the form  (the scaling is presented in Ref.[5])
\begin{equation}
	\label{model}
	C\ddot{x} + \dot{x} = -U'(x) + I(t) + \sqrt{2Q}\,\xi(t).
\end{equation}
The dot and prime denotes a differentiation with respect to the dimensionless time $t$ and the rescaled phase \mbox{$x = (\varphi + \pi)/2$}, where $\varphi = \varphi_u +\varphi_d$. The dimensionless capacitance of the device is equal to $C$. The spatially periodic potential $U(x)$ of the period $2\pi$ is in the following form
\begin{figure}[t]
	\centering
	\includegraphics[width=0.85\linewidth]{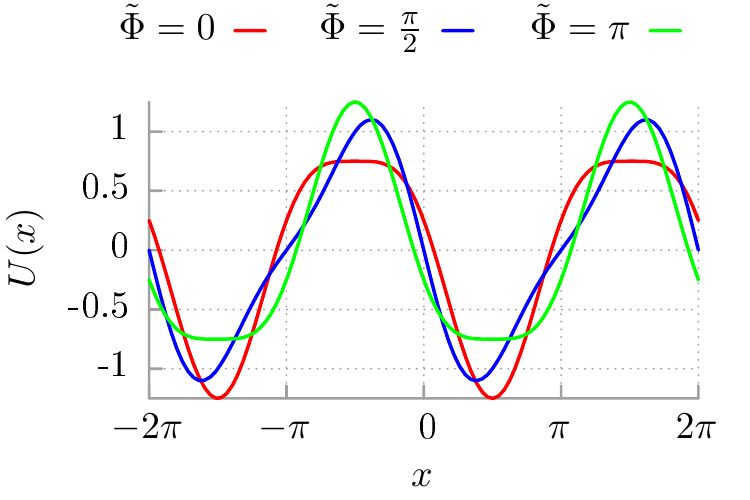}
	\caption{The potential given by Eq. (\ref{potential}) depicted in the symmetric case $\tilde{\Phi} = 0,\pi$ in comparison with the ratchet form $\tilde{\Phi} = \pi/2$.}
	\label{fig2}
\end{figure}
\begin{equation}
	\label{potential}
	U(x) = -\sin{x} - \frac{1}{4}\sin{(2x + \tilde{\Phi} - \pi/2)},
\end{equation}
where $\tilde{\Phi}$ is the dimensionless external constant magnetic flux. When $\tilde{\Phi} \neq 0,\pi$ the reflection symmetry of the potential is broken meaning that there is no such a value of $x_0$ that $U(x - x_0) = U(x + x_0)$ holds for any $x$. Then we classify it as a ratchet profile \cite{hanggi2009}, see Fig. \ref{fig2}. The external driving current $I(t)$ is assumed to be in the harmonic form of the amplitude $a$ and the angular frequency $\omega$, namely, 
\begin{equation}
	\label{driving}
	I(t) = a\cos{(\omega t)}.
\end{equation}
Johnson-Nyquist thermal noise is modelled by symmetric and unbiased $\delta$-correlated white noise $\xi(t)$ of Gaussian nature, i.e., 
\begin{equation}
	\label{noise}
	\langle \xi(t) \rangle = 0, \quad \langle \xi(t)\xi(s) \rangle = \delta(t-s).
\end{equation}
Its intensity $Q \propto k_BT$ is proportional to thermal energy, where $k_B$ and $T$ is the Boltzmann constant and temperature, respectively.

The actual voltage $V(t)$ is proportional to the Josephson phase velocity $V(t) = (\hbar/2e) \dot \varphi(t)$ and the dimensionless voltage is equal to rescaled Josephson phase velocity $v(t) = \dot x(t)$. Due to the presence of the external periodic driving $a\cos{(\omega t)}$  and dissipation the Josephson phase velocity $\dot{x}(t)$ approaches for $t \to \infty$ a unique asymptotic non-equilibrium time-dependent state which is characterized by a temporally periodic probability density  for the Josephson phase velocity $v(t)=\dot{x}(t)$. In consequence, its mean value $\langle \dot{x}(t) \rangle$ in the asymptotic state  for long times ($t \gg 1$) 
takes the form of a Fourier series over all possible harmonics \cite{jung1993}
\begin{equation}
	 \langle \dot{x}(t) \rangle = \langle \mathbf{v} \rangle + v_\mathsf{1T}(t) + v_{2\mathsf{T}}(t) + ...,
\end{equation}
where $\langle \cdot \rangle$ denotes averaging over all realizations of thermal Gaussian white noise as well as  over initial conditions for the Josephson phase $x(0)$ and its velocity $\dot{x}(0)$. Then $\langle \textbf{v} \rangle$ is the dc (time-independent) voltage drop across the SQUID while $v_{n \mathsf{T}}(t)$ are periodic components  of vanishing average over the fundamental period $\mathsf{T} = 2\pi/\omega$. Due to this particular decomposition it is useful to study the period averaged Josephson phase velocity $\mathbf{v}(t)$ defined as
\begin{equation} \label{vT}
	\mathbf{v}(t) = \frac{1}{\mathsf{T}} \int_t^{t + \mathsf{T}} ds\,\dot{x}(s)
\end{equation}
which may be exploited to evaluate the dc voltage $\langle \mathbf{v} \rangle$ 
 from $\langle \mathbf{v}(t) \rangle$ in the long time regime ($t\gg 1$).   A sufficient and necessary condition for the emergence of directed transport $\langle \mathbf{v} \rangle \neq 0$ is breaking of the mirror symmetry of the potential $U(x)$ which is the case for the form given by Eq. (\ref{potential}) \cite{hanggi2009}. This operating principle can be seen as a key for understanding the intracellular transport \cite{bressloff2013}.

We note that Eq. (\ref{model}) is valid only under the assumption that both junctions connected in the series in the bottom arm of the SQUID, see Fig. \ref{fig1}, are identical. In other words it translates to the synchronization of the phase solutions across the junctions and allows to utilize the conservation of supercurrent in this arm which in turn yields Eq. (\ref{model}). The reader may read all the details of the derivation in Ref. [5]. There also the consequences of the assumption of the junctions identity are discussed. The latter is certainly difficult to achieve in reality. However, in this work we will be concerned with the Josephson voltage across the device, i.e. the average behaviour of the effective phase velocity across the SQUID, therefore the assumed synchronized phase approximation is to some extent justified as a guide towards the observable reality rather than taken as the exact solution. Without this assumption the parameter space of the system will be much more complex and difficult to explore even by the numerical means. However, as the first step towards this generalization one could employ e.g. the Ohta semiclassical approach which is exemplified in Ref. [31] and refs. therein. After all, although the employed approximation may be questionable the reader should keep in mind that our results can also be verified in other setups which are free of this assumption, e.g. dynamics of cold atoms in optical lattices. We outlined them in the introductory section of the article.

The above equation (\ref{model}) can be interpreted in the mechanical framework as a classical Brownian particle moving in the periodic potential landscape $U(x)$. Then the particle position $x$ translates to the Josephson phase $\varphi$. Its velocity $v = \dot{x}$ corresponds to the voltage drop $V \propto \dot{\varphi}$ across the SQUID, the particle mass to the capacitance $C$ of the device and the friction coefficient to its conductance. The external force applied to the particle translates then into the current $I(t)$ applied to the SQUID, for details see Ref. [\onlinecite{kautz1996}]. Despite the relative conceptual simplicity of this mechanical picture the analysed system is able to exhibit an extremely rich dynamics and variety of anomalous transport features including the absolute negative mobility \cite{machura2007, slapik2018}, the nonequilibrium noise enhanced transport efficiency \cite{spiechowicz2013jstatmech, spiechowicz2014pre,spiechowicz2016jstatmech},  the anomalous diffusion \cite{metzler2014, spiechowicz2015pre, spiechowicz2016scirep,spiechowicz2017scirep}, the amplification of normal diffusion \cite{reimann2001} and the non-monotonic temperature dependence of normal diffusion \cite{lindner2001, lindner2016, spiechowicz2016njp, spiechowicz2017chaos}. Finally, it is worth to note that the recent progress in fabrication and manufacturing of Josephson junctions allows to construct the Josephson junction ratchet similar to (\ref{model}) using only a single superconducting element \cite{goldobin2013, menditto2016prb, menditto2016}.

The Langevin equation (\ref{model}) cannot be effectively handled by any standard analytical methods. Therefore appropriate numerical simulations have to be be performed in order to analyse dynamics of the  SQUID. Their technical details are described in Ref. [\onlinecite{spiechowicz2015cpc}]. All calculations have been done by use of a CUDA environment implemented on a modern desktop GPU. This proceeding allowed for a speedup of a factor of the order $10^3$ times as compared to a common present-day CPU method \cite{spiechowicz2015cpc}. Various properties of the system (\ref{model}) have been presented in a series of our previous papers \cite{spiechowicz2014prb,spiechowicz2015njp, spiechowicz2015chaos,spiechowicz2015pre,spiechowicz2016scirep, spiechowicz2017chaos}. Here, we analyse another aspect: statistics of transitions for an approximate discrete three-state stochastic process which mimics three deterministically coexisting attractors for the voltage in the regime of thermally induced dynamical localization of the Josephson phase velocity.
\begin{figure}[t]
	\centering
	\includegraphics[width=0.85\linewidth]{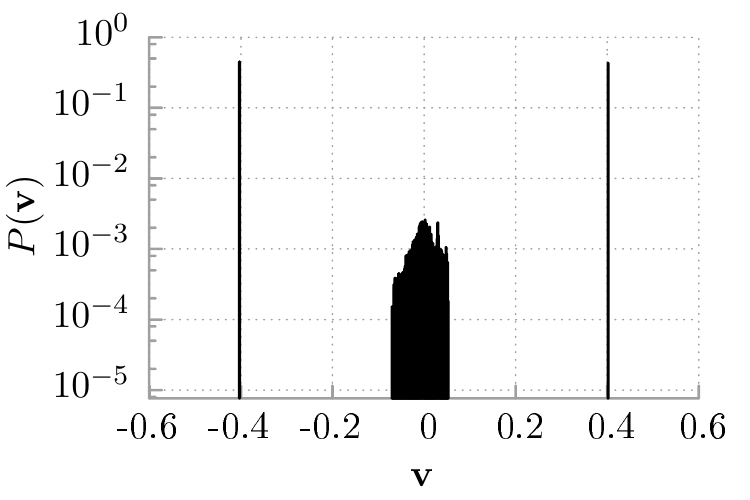}
	\caption{The probability distribution $P(\mathbf{v}(t))$ of the asymptotic long time period averaged Josephson phase velocity $\mathbf{v}(t)$  is presented for the deterministic system $Q=0$ with $C=6$,  \mbox{$a = 1.899$}, $\omega = 0.403$ and $\tilde{\Phi} = \pi/2$.}
	\label{fig3}
\end{figure}
\subsection{Deterministic dynamics: Q=0}
The phase space $\{x,\dot{x},\omega t\}$ of the noiseless autonomous system  equivalent to (\ref{model}) is three
-dimensional (which is minimal for it to display chaotic evolution) and the parameter space $\{m, a, \omega, Q, \tilde{\Phi}\}$  is five-dimensional leading to very rich and complex features of the system \cite{mateos2000}. Its analysis is  by no means trivial. Systematic numerical investigation of this system is a challenging task even for modern computers. 
Therefore to reduce a number of free parameters of the model we will first consider only a special case  $\tilde{\Phi} = \pi/2$ for which the asymmetry is the most pronounced, see Fig. \ref{fig2}. Furthermore, since in many cases noise assisted dynamics may be a consequence of the corresponding deterministic evolution as the starting step we analyse the noiseless case $Q = 0$. 
In the following, we consider a limited part of the parameter space: \mbox{$C=6, a=1.899, \omega=0.403$ and $\tilde{\Phi}=\pi/2$}. It is a very interesting regime of transient anomalous diffusion \cite{spiechowicz2015pre}. For this particular set of parameters the system is {\it non-chaotic}. By performing precise numerical simulations of the studied system we found that there are two running states corresponding to either positive $v_+ = \mathbf{v}(t) \approx 0.4$ or negative $v_- = \mathbf{v}(t) \approx -0.4$ period averaged voltage and the locked state for which it is negligibly small $v_0 = \mathbf{v}(t) \ll 1$. This family of solutions forms a set of three attractors $\{v_+, v_-, v_0\}$.  Each attractor $v_k \,(k=+,-,0)$  has its own domain of attraction $\mathcal{D}_k$ \cite{spiechowicz2016scirep}. These are disjoint, $\mathcal{D}_k \cap \mathcal{D}_{k'} = \emptyset$ for $k\ne k'$,  and partition the total available state space. Therefore the dynamics of the system is {\it non-ergodic}. We exemplify this typical structure of solutions in Fig. \ref{fig3},  where we depict the probability distribution $P(\mathbf{v}(t))$  in the long time regime. 
\begin{figure}[t]
	\centering
	\includegraphics[width=0.85\linewidth]{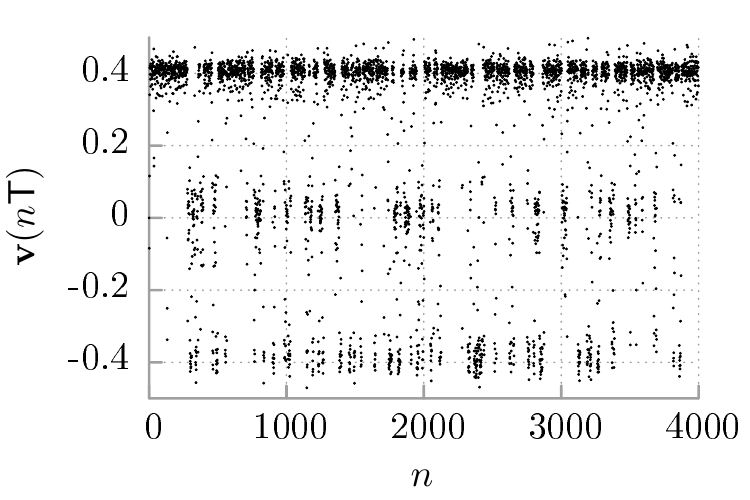}
	\caption{The exemplary trajectory of period averaged Josephson phase velocity $\mathbf{v}(n\mathsf{T}) = \mathbf{v}(t=n\mathsf{T})$ is presented for noisy system $Q = 0.005$. Other parameters are the same as in Fig. \ref{fig3}.}
	\label{fig33}
\end{figure}
\section{Statistics of transitions}  
For non-zero noise intensity $Q>0$ thermal noise activates the stochastic dynamics which enables random transitions between coexisting deterministic disjoint attractors $\{v_+, v_-, v_0\}$. We now analyse statistics of transitions between these states. 
For this purpose we roughly estimate the continuous Langevin dynamics (\ref{model}) with the discrete three state stochastic process $\{v_+, v_-, v_0\}$ with jumps between the states induced by thermal equilibrium fluctuations $\xi(t)$ and the external forcing $I(t)$. These transitions can be approximately analysed in terms of the Markov processes. There are two ways to solve this problem: either as a time-continuous Markov process or as a Markov chain. The second option is adequate in our case since we deal with the sequences of the period averaged Josephson phase velocity $ \mathbf{v}(n\mathsf{T}) = \mathbf{v}(t=n\mathsf{T})$, cf. Eq. (\ref{vT}) and  Fig. \ref{fig33}. The transition matrix $\mathbf{M}$ describing this Markov chain has the form
\begin{equation}
\mathbf{M}= \left( \begin{array}{ccc}
p_{++} & p_{+0} & p_{+-}\\
p_{0+} & p_{00} & p_{0-}\\
p_{-+} & p_{-0} & p_{--}
\end{array}
\right),
\quad \sum _j p_{kj} =1, 
\label{M}
\end{equation}
where $p_{++}$ stands for the conditional probability to remain staying in the plus state $v_+ \to v_+$, $p_{+0}$ is the probability of a transition between the plus and the zero state $v_+ \to v_0$ and $p_{+-}$ denotes the probability for a jump between the opposite states $v_+ \to v_-$.  This convention is analogous for the remaining six probabilities $\{p_{00}, p_{0+}, p_{0-}, p_{--}, p_{-0}, p_{-+}\}$. By introducing the threshold $\nu = 0.2$ the above quantities can be calculated from simulations of the Langevin dynamics (\ref{model}) as the relative frequencies with which the period averaged Josephson phase velocity $\mathbf{v}(t)$ visits the three coarse grained regions $V_+ = \{|\mathbf{v}(t) - 0.4| < \nu \}$, \mbox{$V_0 = \{|\mathbf{v}(t)| < \nu \}$} and $V_- = \{|\mathbf{v}(t) + 0.4| < \nu \}$. A chosen value of the threshold $\nu = 0.2$ follows naturally from inspection of the probability distribution of the period averaged Josephson phase velocity where for low to moderate temperature regimes there are three peaks corresponding to the deterministic coexisting attractors which are approximately $\nu = 0.2$ apart, see Fig. 3 for zero and Fig. 4 for high temperature. To determine transition probabilities we averaged over $10^5$ numerically calculated system trajectories each lasting for $10^6$ periods of the external ac driving $\mathsf{T} = 2\pi/\omega$ and starting from different initial condition $(x(0),\dot{x}(0))$ taken randomly from the interval $[0,2\pi] \times [-2,2]$.
\begin{figure}[t]
	\centering
	\includegraphics[width=0.85\linewidth]{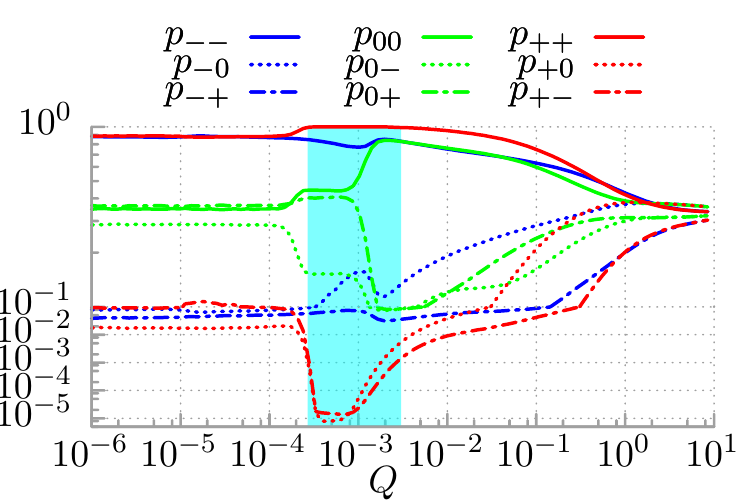}
	\caption{All transition probabilities between the three observed states: the minus $v_- = -0.4$, the zero $v_0 = 0$ and the plus $v_+$ solution depicted versus thermal noise intensity $Q$. Other parameters are the same as in Fig. \ref{fig3}.}
	\label{fig4}
\end{figure}
\subsection{Noise induced dynamical localization}
In Fig. \ref{fig4} we present all transition probabilities between the observed three states:  minus $v_- = -0.4$,  zero $v_0 = 0$ and plus $v_+$ solutions depicted in the long time limit versus thermal noise intensity $Q$. For low temperature regime we observe that both the probabilities $p_{--}$ and $p_{++}$ for the Josephson phase velocity to survive in the $v_-$ and $v_+$ state, respectively, are significantly larger than others. It is worth to note that then the magnitude of $p_{--}$ and $p_{++}$ is almost equal and not so far from one. In contrast, the probability for staying in the locked state $p_{00}$ is twice lower. With the cyan colour we marked the region of moderate temperature where thermal noise induces dynamical localization \cite{guarneri2014, paul2016, bitter2017, notarnicola2017} of the period averaged Josephson phase velocity (voltage), i.e. it resides in the plus state with the probability $p_{++}$ very, very close to unity, namely  $p_{++} = 0.9999992$. In terms of the mechanical model, it means that once the ensemble of particles enters this state, particles move almost coherently with the same velocity and with marginal fluctuations allowed by the chosen threshold $|\mathbf{v}(t) - 0.4| < \nu$. In this interval of temperature the survival probability in the locked state $p_{00}$ increases while the corresponding quantity for the minus state $p_{--}$ passes through its local minimum. Further grow of thermal noise intensity immediately delocalizes the Josephson phase velocity, $p_{++}$ decreases and all transition probabilities progressively converge to the same value. In the limit of high temperature they all become equivalent.
\begin{figure}[t]
	\centering
	\includegraphics[width=0.85\linewidth]{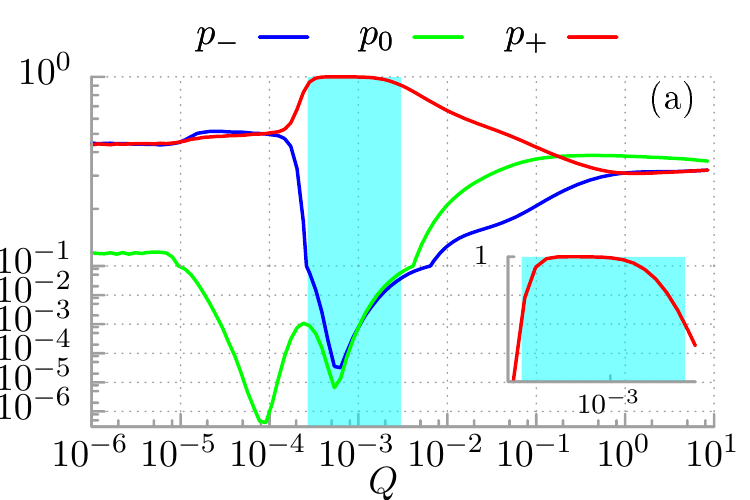}\\
	\includegraphics[width=0.85\linewidth]{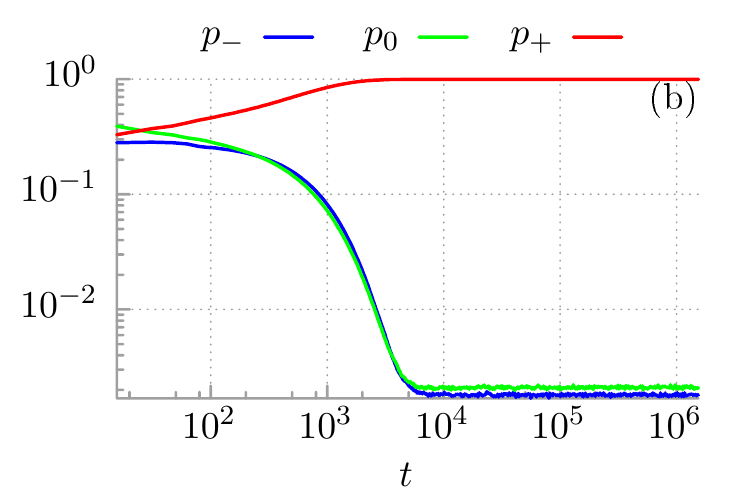}
	\caption{Panel (a): The probability $p_-$, $p_0$ and $p_+$ for the phase velocity to be in the state $v_-$, $v_0$, $v_+$, respectively presented as a function of thermal noise intensity $Q$. Panel (b): time evolution of the above probabilities for temperature $Q = 0.001175$. Other parameters are the same as in Fig. \ref{fig3}.}
	\label{fig5}
\end{figure}
\begin{figure}[b]
	\centering
	\includegraphics[width=0.85\linewidth]{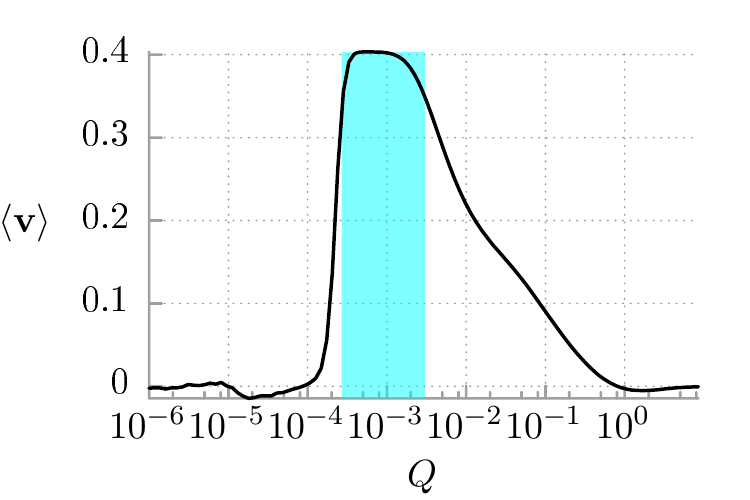}
	\caption{The dc (time-independent) voltage drop $\langle \mathbf{v} \rangle$ across the SQUID versus thermal noise intensity $Q$. Other parameters are the same as in Fig. \ref{fig3}.}
	\label{fig6}
\end{figure}
\begin{figure*}[t]
	\centering
	\includegraphics[width=0.3\linewidth]{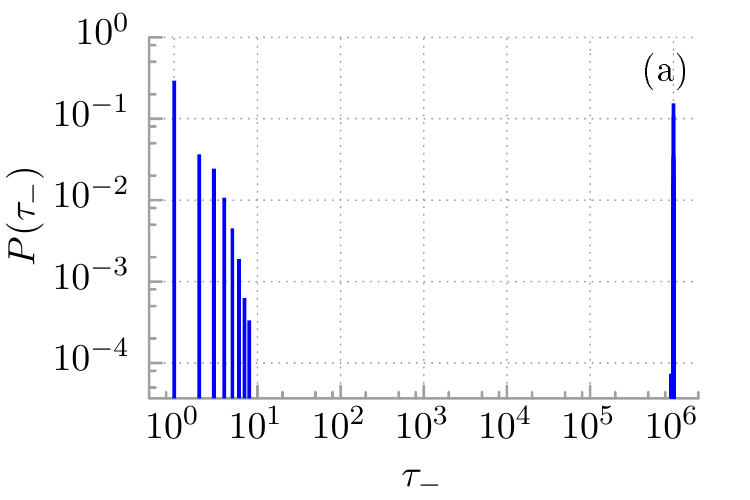}
	\includegraphics[width=0.3\linewidth]{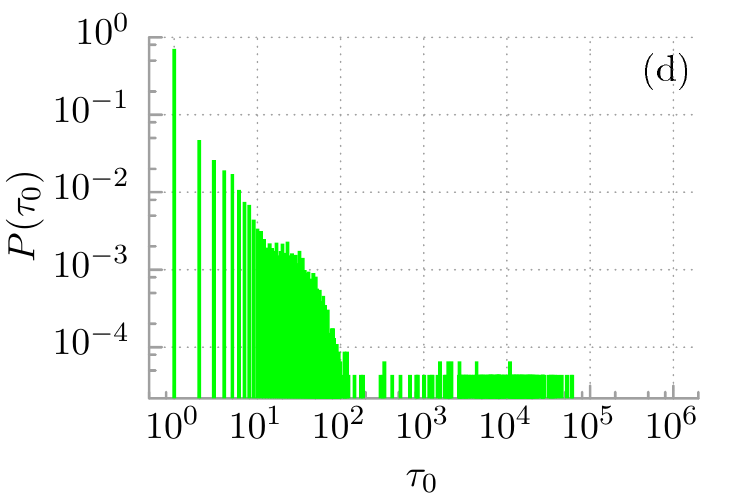}
	\includegraphics[width=0.3\linewidth]{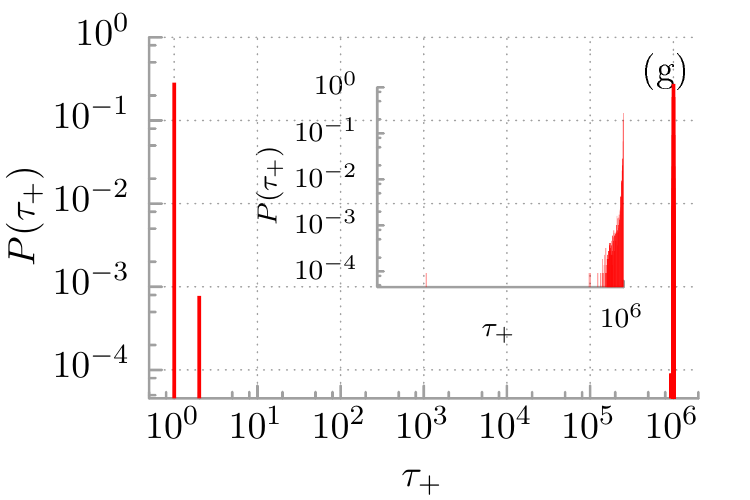}\\
	\includegraphics[width=0.3\linewidth]{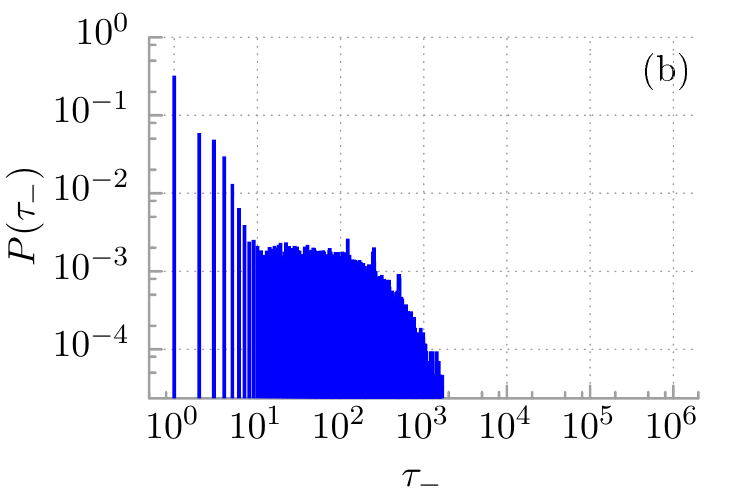}
	\includegraphics[width=0.3\linewidth]{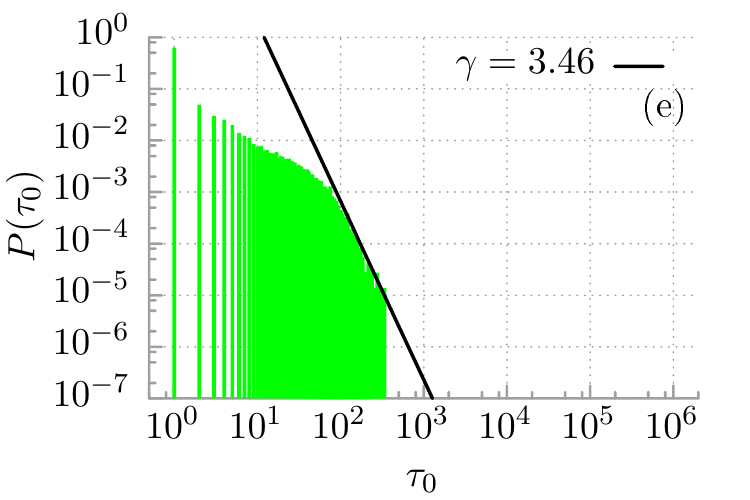}
	\includegraphics[width=0.3\linewidth]{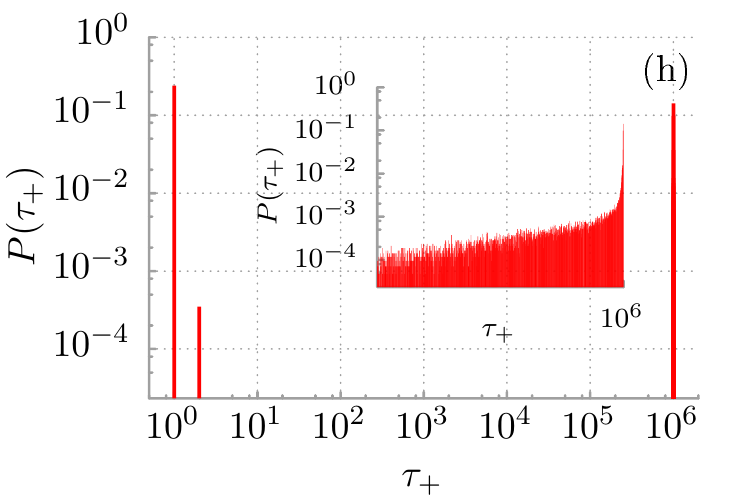}\\
	\includegraphics[width=0.3\linewidth]{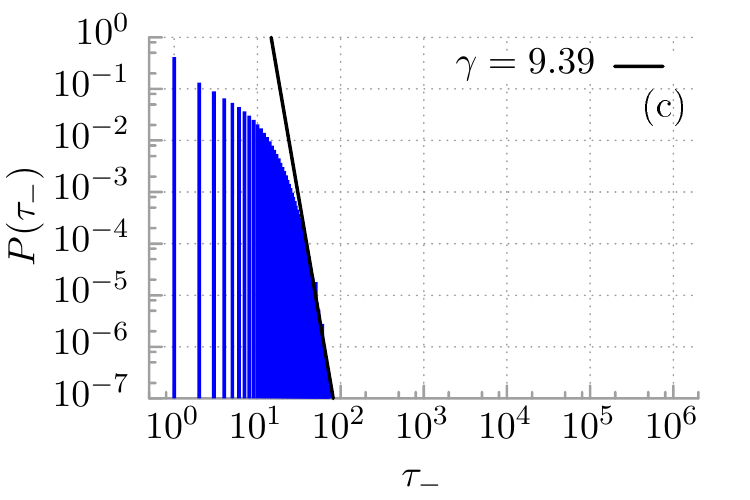}
	\includegraphics[width=0.3\linewidth]{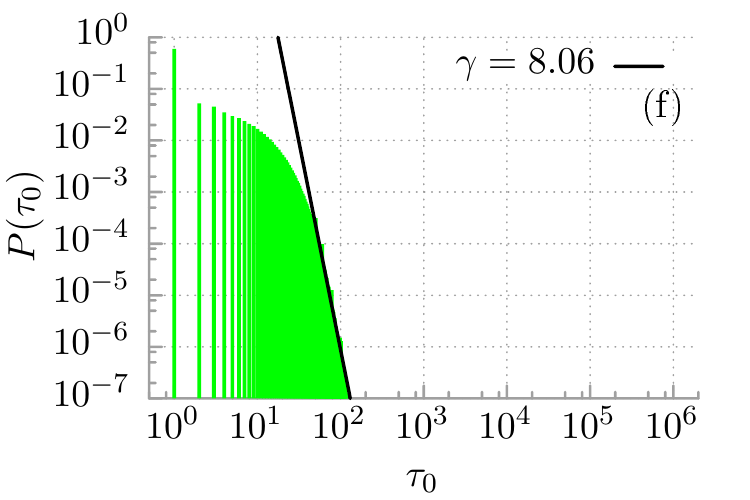}
	\includegraphics[width=0.3\linewidth]{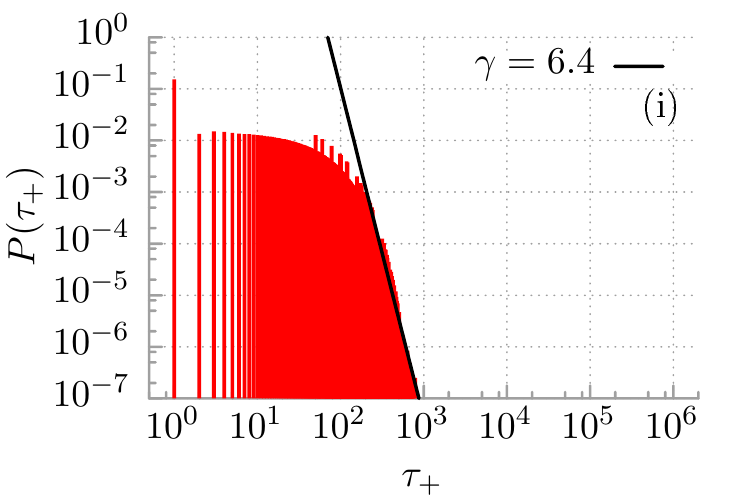}
	\caption{Residence time probability distribution $P(\tau_-)$ (panel (a)-(c)), $P(\tau_0)$ (panel ((d)-(f)) and $P(\tau_+)$ (panel (g)-(i)) for the states $v_-$, $v_0$ and $v_+$, respectively, is shown for different thermal noise intensity $Q = 10^{-5}$ (upper row), $Q = 0.0004$ (middle row) and $Q = 0.005$ (bottom row). For other parameters see Fig. \ref{fig3}}
	\label{fig9}
\end{figure*}

Let us now answer the question about the fraction of the Josephson phase velocities (voltage) which get localised in the plus state. For this purpose we additionally consider the probabilities $p_-$, $p_0$ and $p_+$ for the Josephson phase velocity to be in the state $v_-$, $v_0$ and $v_+$ in the long time limit, respectively. These quantities are shown in Fig. \ref{fig5} (a) as a function of thermal noise intensity $Q$. For low temperature regime almost equal fraction of Josephson phase velocities is observed in the minus and the plus state. At the same time the group of the locked states $v_0$ is several times smaller. This situation is drastically changed for moderate temperature where, as we suggested before, the Josephson phase velocity (voltage) gets localised. Then the fraction of the minus and zero states rapidly decreases up to temperature $Q \approx 4 \times 10^{-4}$ where they both are minimal and negligibly small. Consequently, almost all velocities are localized in the plus solution. Moreover, a careful inspection of this panel reveals that the range of temperature where the probability for the Josephson phase velocity to survive in the plus state $p_{++}$ is equal almost one does not exactly overlap with the corresponding interval where all velocities resides in this state which is marked by the cyan colour, see the inset of Fig. \ref{fig5} (a). For further increase of thermal noise intensity the probabilities $p_-$, $p_0$ and $p_+$ approach each other and finally they assume the same value in the high temperature limit. In  panel (b) of the same figure we depict time evolution of the probabilities $p_-$, $p_0$ and $p_+$ for the Josephson phase velocity to be in the plus, zero and minus states, respectively. As exemplified, for temperature taken from the range where thermal noise intensity induces dynamical localization of the Josephson phase velocity the probability $p_+$ is a non-decreasing function of time. In other words it means that the mentioned dynamical localization is not a transient effect but rather a persistent phenomenon. We confirmed this conjecture for other temperatures and longer time-scales of the Langevin dynamics simulations (not depicted).
\begin{figure}[t]
	\centering
	\includegraphics[width=0.85\linewidth]{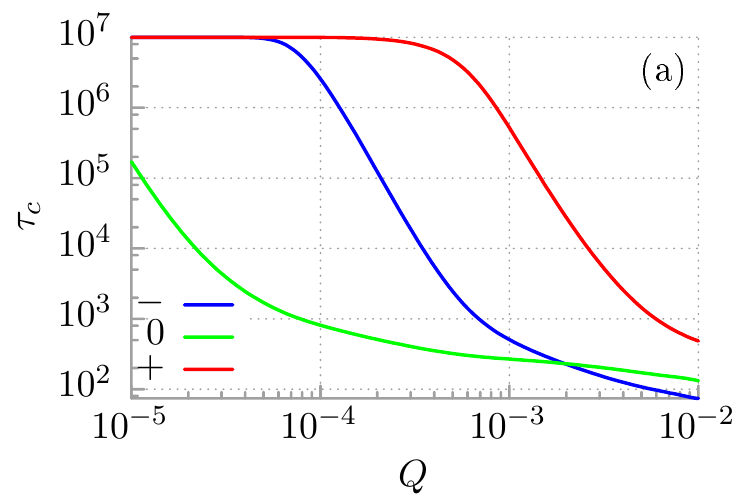}\\
	\includegraphics[width=0.85\linewidth]{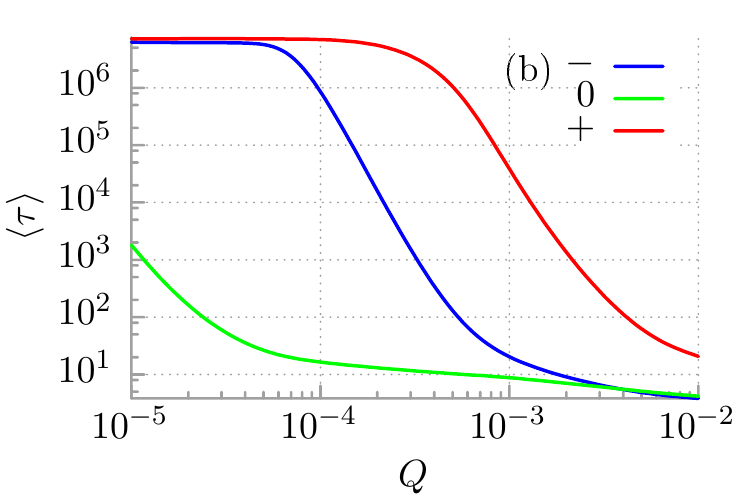}
	\caption{Panel (a): the cut-off residence time $\tau_c$ as a function of temperature of the system $Q$ for the state $v_-$, $v_0$ and $v_+$. Panel (b): the mean value of the residence time $\langle \tau \rangle$ depicted versus the same quantity. Other parameters are the same as in Fig. \ref{fig3}.}
	\label{fig10}
\end{figure}
\subsection{Impact on the dc voltage}
Noise induced transitions between coexisting attractors naturally influence the directed transport expressed as the dc (time-independent) voltage drop $\langle \mathbf{v} \rangle$ across the studied SQUID setup. In Fig. \ref{fig6} we present the dependence of the dc voltage drop $\langle \mathbf{v} \rangle$ on temperature $Q \propto T$. For vanishing thermal noise intensity $Q \to 0$ the directed transport is negligibly small $\langle \mathbf{v} \rangle = 0$ which is in agreement with the probability distribution of the individual asymptotic long time period averaged Josephson phase velocity $P(\mathbf{v}(t))$ shown in Fig. \ref{fig3} for the deterministic variant of the system (\ref{model}) with $Q = 0$. It is so because of the fact that simply the weighted average over the two running attractors $v_- = 0.4$ and $v_+ = 0.4$ as well as the locked state $v_0 = 0$ gives zero $\langle \mathbf{v} \rangle = 0$. For slightly higher temperature we observe small fluctuations around the deterministic value $\langle \mathbf{v} \rangle = 0$ up to the moment where the dynamical localization of the Josephson phase velocity occurs. Again, for the convenience of the reader we marked this region with the cyan colour. We note there the so called thermal noise induced ratchet effect \cite{spiechowicz2014prb}, i.e. the thermally activated emergence of the directed transport in the nonequilibrium system with broken spatiotemporal symmetry. By analysing the dynamics of  random transitions between certain regions in the phase space of the system we are actually able to understand the origin of this effect from the point of view of microscopic dynamics occurring in the phase space. It is solely rooted in the dynamical localization of the Josephson phase velocity and its temperature robustness which is illustrated in Fig. \ref{fig4}. In particular, the magnitude of plateau of the dc voltage $\langle \mathbf{v} \rangle = 0.4$ in Fig. \ref{fig6} is a direct consequence of the localization in the plus state $v_+ = 0.4$. Moreover, the temperature range where the plateau is visible matches exactly the region where all phase velocities get localised in the plus state $p_+ = 1$, c.f. the inset of Fig. \ref{fig5} (a). Higher temperature immediately destroys the dynamical localization and consequently causes the gradual decrease of the dc voltage drop $\langle \mathbf{v} \rangle$ across the device. In high temperature limit the probabilities for the Josephson phase velocity to be in the minus, the plus and the zero state becomes almost equivalent which as a result leads to vanishing of the dc voltage $\langle \mathbf{v} \rangle = 0$.

It is worth to note that the discussed thermal noise induced dynamical localization manifests itself also as subdiffusion of the Josephson phase. We refer the interested reader to our recent work on this curious problem, see Ref. [\onlinecite{spiechowicz2017scirep}].

\begin{figure}[t]
	\centering
	\includegraphics[width=0.85\linewidth]{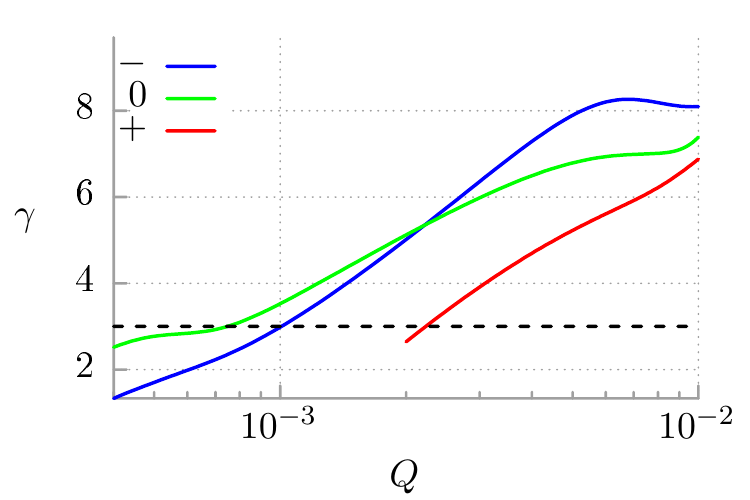}
	\caption{Temperature dependence of the power law tails exponent $\gamma$ of the residence time distributions $P(\tau)$ is shown for the states $v_-$, $v_0$ and $v_+$ in the interval where the broad tails of the distributions can be well fitted in this way. Other parameters are the same as in Fig. \ref{fig3}.}
	\label{fig11}
\end{figure}
\subsection{Power law tails in the residence time probability distributions}
The approximation of the continuous Langevin dynamics by a discrete stochastic process allows us to study the residence time distribution in the states $v_-$, $v_0$ and $v_+$. We define the residence time $\tau$ as a  number of temporal periods $\mathsf{T} = 2\pi/\omega$ of the external driving $a\cos{(\omega t)}$ which the Josephson phase velocity waits in the given state until it jumps to the other one. In Fig. \ref{fig9} we depict the probability distribution $P(\tau)$ of the residence time 
in  the states  $v_-$, $v_0$ and $v_+$ calculated from an ensemble of $10^5$ trajectories which evolved up to the final moment of time $t = 10^6 \mathsf{T}$.    Upper panels corresponds to low temperature regime of thermal noise intensity $Q = 10^{-5}$. Bimodal distribution is then observed for the state $v_+$ and $v_-$. The Josephson phase velocity either jumps out of the state after several periods of the external driving or stays there for nearly the whole duration of simulations. We stress that the peak corresponding to the total duration of the simulation $\tau = 10^6$ means that the Josephson phase velocity resides in a  given state for the entire simulation. We present it to ensure correct normalization of the depicted residence time probability distribution. Significantly longer time scales are practically difficult to achieve due to technical restrictions of the presently accessible computer hardware. We note that for the limiting case of vanishing thermal noise intensity $Q \to 0$ the Josephson phase velocity  may reside in these states arbitrary long. The residence time distribution $P(\tau_0)$ for the zero state $v_0$ is unimodal but possesses the broad tail. Most likely the Josephson phase velocity will quickly jump out of this state onto the minus or plus solution. The situation changes drastically for low to moderate temperature regime where the dynamical localization of the Josephson phase velocity occurs. We exemplify it in the middle row of panels where thermal noise intensity is $Q = 0.0004$. Now only the residence time distribution $P(\tau_+)$ for the plus state displays the bimodal character. The previously well pronounced peak corresponding to $\tau_- = 10^6$ in the minus distribution $P(\tau_-)$ no longer exists and is replaced with the broad tail. These features implies the mentioned noise induced dynamical localization of the Josephson phase velocity. The impact of intensity of thermal fluctuations on the peak corresponding $\tau_+ = 10^{6}$ in $P(\tau_+)$ is illustrated in insets of panels Fig. \ref{fig9} (g) and Fig. \ref{fig9} (h). When temperature is increasing this peak is generally flattened and at the same time broadened towards lower values of $\tau_+$. Finally, the moderate to high temperature regime for $Q = 0.005$ is depicted in the bottom row of panels. Then the bimodality of the residence time distribution is destroyed by thermal fluctuations even for the plus state $v_+$. All distributions look similar,  however they slightly differ in the asymptotic parts. To quantify this aspect we successfully fitted to them a power law tail in the form $P(\tau) = \tau^{-\gamma}$. For sufficiently high thermal noise intensity $Q$ the temperature dependent exponents are greater than three,  $\gamma > 3$,  which according to Ref. [\onlinecite{weeks1998}] indicates normal diffusion of the Josephson phase. This agrees well with our intuition since in the high temperature limit thermal noise dominates the dynamics and other contributions become negligible.

In Fig. \ref{fig10} (a) we present the cut-off residence time $\tau_c$ as a function of temperature of the system $Q$ for the state $v_-$, $v_0$ and $v_+$. We define this quantity as the largest time $\tau$ such that $P(\tau) \neq 0$. We note that the finiteness of $\tau_c$ for the plus $v_+$ and minus $v_-$ solution in the limit of vanishing thermal noise $Q \to 0$ is due to the technical restriction of the total simulation time. Normally it will diverge to infinity $\tau_c \to \infty$ reflecting the coexistence of the deterministic disjoint attractors for $v_+$ and $v_-$. However, the finiteness of $\tau_c$ for the noisy system is not a technical artefact but rather its expected feature. The cut-off times for all the observed states are monotonically decreasing functions of temperature. Their magnitude are related to the crossover times of transient anomalous diffusion occurring in this setup \cite{spiechowicz2017scirep}. For example, the cut-off time $\tau_c$ for the minus state $v_-$ is related to the duration of superdiffusion whereas the corresponding quantity for the plus solution $v_+$ can be associated with the total period of subdiffusion \cite{spiechowicz2017scirep}. In panel (b) of the same figure we additionally present the mean residence time $\langle \tau \rangle$ as a function of thermal noise intensity $Q$ for all three states. It looks very similar to the curve of the cut-off time $\tau_c$ thus indicating the prominent role of the broad tail in observed distributions. For times much larger than the largest cut-off time $t \gg \tau_c$ normal diffusion is expected to be observed.

Finally, in Fig. \ref{fig11} we present the temperature dependence of the power law tails exponent $\gamma$ of the residence time distribution $P(\tau)$ for the three states in the interval where the broad tails of the distributions can be well fitted in this way. The dashed horizontal line indicates $\gamma = 3$ which is critical for the emergence of the normal diffusion. Accordingly, for temperature $Q < 0.002$ at the time scale of the order of largest cut-off residence time $\tau_c$ various regimes of transient anomalous diffusion are present which will eventually give place to normal diffusion \cite{spiechowicz2016scirep,spiechowicz2017scirep}. On the other hand, for $Q > 0.002$ normal diffusion is expected to be found even for times comparable to the largest cut-off residence time $\tau_c$.
\begin{figure}[t]
	\centering
	\includegraphics[width=0.85\linewidth]{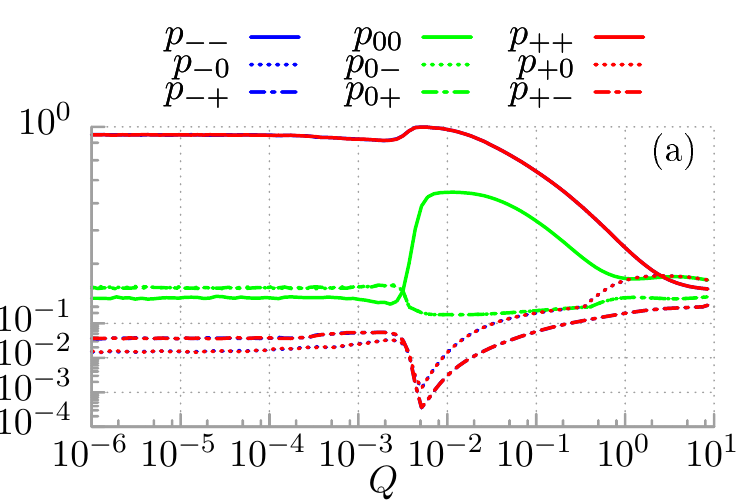}\\
	\includegraphics[width=0.85\linewidth]{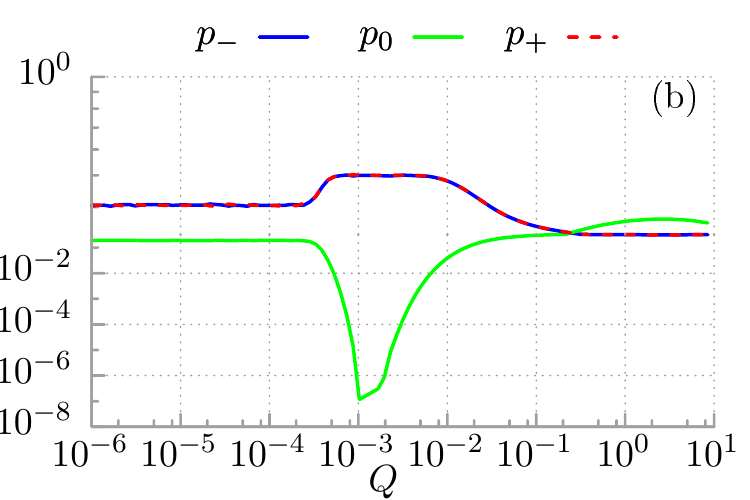}
	\caption{Panel (a): all transition probabilities between the three observed states $v_-$, $v_0$ and $v_+$. Panel (b): the probabilities $p_-$, $p_0$ and $p_+$ for the phase velocity to be in the above states. All characteristics are shown versus thermal noise intensity $Q$ and for the regime corresponding to parameters listed in Fig. \ref{fig3} except the fact that now the system is symmetric with $\Phi = \pi$, c.f. Fig. \ref{fig2}. Due to the latter fact the red and blue lines overlap in plots.}
	\label{fig8}
\end{figure}
\subsection{The role of symmetry breaking}
Natural question which at some point comes to mind is whether the behaviour similar to the considered dynamical localization may be observed in unbiased symmetric systems. Then due to the symmetry a preferential direction of motion is impossible in the stationary state and $\langle \mathbf{v} \rangle \equiv 0$ \, \cite{denisov2014}.  To answer this question we consider the case \mbox{$\tilde{\Phi} = \pi$} for which the spatially periodic potential $U(x)$ is symmetric, c.f. Fig. \ref{fig2}. In the deterministic case $Q = 0$ and for exactly the same set of the remaining model parameters, see Fig. \ref{fig3}, the system is non-ergodic and possesses two regular attractors transporting in the opposite direction with the same magnitude $v_+ = - v_- = 0.4$ as well as one locked solution $v_0=0$, see addendum in Ref. [\onlinecite{spiechowicz2015pre}]. In Fig. \ref{fig8} we present all transition probabilities between the observed three states $v_-$, $v_0$ and $v_+$ as well as the corresponding probabilities for the Josephson phase velocity to be in these solutions $p_-$, $p_0$ an $p_+$ versus thermal noise intensity $Q$. 
 
When $Q>0$ the system is ergodic and statistical properties of the system can be deduced from one sufficiently long trajectory. It means that it cannot be localized in one non-zero state, say with $v>0$, for an arbitrary long time because the symmetry implies $\langle \mathbf{v} \rangle = 0$. This is a fundamental difference that distinguishes symmetric and asymmetric systems. In the latter the dynamical localization may be in principle a persistent effect. However, a pair of solutions $v_{-\varepsilon} = -\varepsilon$ and $v_{\varepsilon} = \varepsilon$ may exist in symmetric system dynamics. Then trajectories can be simultaneously localized in these states meaning that $p_{-\varepsilon-\varepsilon} = 1$ and $p_{\varepsilon\varepsilon} = 1$. The latter will eventually result in condition $p_{-\varepsilon} = p_{\varepsilon}$ which must be obeyed in order to satisfy $\langle \mathbf{v} \rangle = 0$. We exemplify this scenario in Fig. \ref{fig8}.

\section{Conclusion}
With this work we studied the dynamics of occupations of certain regions of the phase space of  the asymmetric SQUID device which is driven by the external harmonic current and pierced by the external magnetic flux. 
In the mechanical framework it describes a driven Brownian particle moving in the periodic potential of the controllable reflection symmetry.

In the considered set of parameters the deterministic counterpart of the system is in a non-chaotic regime and possesses three coexisting attractors for the Josephson phase  velocity (voltage) averaged over the period of the external driving current: two of them are running solutions pointing to the opposite direction \mbox{$v_\pm = \pm 0.4$} and another one is the locked state $v_0 = 0$. Based on this fact we analysed  transitions between the running and/or locked states in the presence of stochastic dynamics activated by thermal fluctuations. The main message of the paper is the occurrence of the thermal fluctuations induced dynamical localization of the Josephson phase velocity (voltage), i.e. there is a temperature range for which the probability for it to survive in the positive running state $v_+$ is very, very close to one. By considering additionally the probability for the Josephson phase velocity to be in each of the state we demonstrated that it does not necessarily involve that all velocities are found in the $v_+$ solution. Moreover, we discussed implications of this phenomenon on the directed transport (dc voltage) across this device and pointed out its impact on the Josephson phase diffusion in this setup. It allows us to explain the thermal noise induced ratchet effect observed in the dependence of the dc voltage across the device on temperature. We studied also the residence time probability distributions for the state $v_-$, $v_0$ and $v_+$ which confirm our previous findings. In particular, surprisingly in moderate to high temperature regime they possess the broad tails which can be well approximated by the power-law. The cut-off residence times of these distributions are related to the duration of transient anomalous diffusion processes occurring in this setup. Finally, we presented a comment on the role of the symmetry breaking in the thermal noise induced dynamical localizations to show that under certain circumstances it may occur also in driven symmetric and periodic systems.

Dynamical localization is a well recognized effect in quantum systems which has been known for many years \cite{casati1979}. It manifests itself in quantum suppression of chaotic classical diffusion in momentum space due to interference effects \cite{paul2016, bitter2017, notarnicola2017}. With this work we established a relation between the thermally induced dynamical localization in the velocity space of driven periodic systems and their directed transport. Noise induced dynamical localization demonstrates how deterministic-like behaviour may be extracted \emph{out of randomness} itself. After effects like stochastic \cite{vulpiani1981, gammaitoni1998} or coherence \citep{pikovsky1994, lindner2004} resonance it reveals another fascinating face of fluctuations. In view of the widespread presence of periodic systems with or without broken reflection symmetry and across the macro and microscales, our research may contribute to further understanding of especially microworld which is \emph{in situ} immersed in an unavoidable sea of fluctuations.

\section*{Acknowledgement}
J. S. was supported by the Foundation for Polish Science (FNP) START fellowship and J. {\L}. by the Grant NCN 2015/19/B/ST2/02856.

\end{document}